\documentclass[useAMS,usenatbib]{mn2e} 
\usepackage{graphicx}

\newcommand{\be}{\begin{equation}}
\newcommand{\ee}{\end{equation}}
\newcommand{\ba}{\begin{eqnarray}}
\newcommand{\ea}{\end{eqnarray}}

\title[SDSS-GALEX viewpoint of the $z\sim0$ red sequence]{The SDSS-GALEX viewpoint of the truncated red sequence in field
environments at $\bmath{z\!\sim\!0}$}

\author[Haines et al.]{C. P. Haines,$^{1}$ A. Gargiulo$^{1,2}$ and P. Merluzzi$^{1}$ \\
$^{1}$INAF - Osservatorio Astronomico di Capodimonte,
via Moiariello 16, I-80131 Napoli, Italy; chris@na.astro.it \\
$^{2}$Department of Physics, Universit\`{a} Federico II, Napoli, Italy}

\begin{document}

\maketitle
\label{firstpage}
\begin{abstract}
We combine {\em GALEX} near-UV photometry with a volume-limited sample
of local \mbox{($0.005\!<\!z\!<\!0.037$)} SDSS DR4 galaxies to examine the composition and the environmental
dependencies of the optical and UV-optical colour-magnitude (C-M) diagrams. 
We find that \mbox{$\sim\!3$0\%} of red sequence galaxies in the optical C-M
diagram show
signs of ongoing star-formation from their spectra having
\mbox{EW(H$\alpha)>2$\AA}. This contamination is greatest at faint magnitudes \mbox{(M$_{r}\!>\!-19$)} and in field regions where as many as three-quarters of red sequence
galaxies are star-forming, and as such has important consequences for
following the build-up of the red sequence. We find that the
${\rm NUV}-r$ colour instead allows a much more
robust separation of passively-evolving and star-forming galaxies,
which allows the build-up of the UV-selected red sequence with
redshift and environment to be directly interpreted in terms of the assembly of
stellar mass in passively-evolving galaxies. We find that in isolated
field regions the
number density of UV-optical red sequence galaxies declines rapidly at
magnitudes fainter than \mbox{M$_{r}\!\sim\!-19$} and appears completely
truncated at \mbox{M$_{r}\!\sim\!-18$}. This confirms the findings of
Haines et al. (2007) that no passively-evolving dwarf galaxies are
found more than two virial radii from a massive halo, whether that be
a group, cluster or massive galaxy. These results support
the downsizing paradigm whereby the red sequence is assembled from the
top down, being already largely in place at the bright end by
\mbox{$z\!\sim\!1$}, and the faint end filled in at later epochs in
clusters and groups through environment-related processes such as
ram-pressure stripping or galaxy harassment.

\end{abstract}


\section{Introduction}
\label{intro}

The most widely studied bimodality in the galaxy properties is that
observed for their colours, that is the clear separation of galaxies into
the red sequence and blue cloud populations \citep[e.g.][]{strateva}. 
This has the advantage
that it is easy to measure, particularly at high-redshifts, allowing
studies to follow the evolution of the bimodality to \mbox{$z\sim1.2$}
\citep{bell,willmer}, and demonstrating its
existence at even \mbox{$z\sim2$} \citep{giallongo,cirasulo}. The main drawback
of using optical colours is that they do not necessarily fully relate to the
underlying star-formation history, in particular galaxies can appear
red not only because they are passive, but also through high levels of
dust extinction produced by star-bursts. 

A number of studies have followed separately the evolution of the red
and blue galaxy luminosity functions to \mbox{$z\ga\!1$}, in particular with
regard to characterising the build-up of the red sequence
population. There is little evolution in the number
density of massive red sequence galaxies with \mbox{$\mathcal{M}>10^{11}{\rm
  M}_{\sun}$}, with 
only a factor two increase in the stellar mass density from \mbox{$z\sim1$}
to the present day \citep{bell, willmer}, or a factor six since
\mbox{$z\sim2$} \citep{glazebrook}. 
At lower stellar masses however a much more rapid evolution is
observed. By \mbox{$z\sim0.8$} the stellar mass density of red galaxies with
\mbox{$\mathcal{M}\la\!10^{10.3}$} has dropped by an order of magnitude, while
by \mbox{$z\sim1.2$} this cut-off mass has risen to
\mbox{$\mathcal{M}\sim10^{10.6}$} \citep{bundy}.  
The red sequence thus appears to be built up from the top down, with
the most massive galaxies in place first at \mbox{$z>1$}, and the faint end
incrementally filled in at lower redshifts.

The epoch and rapidity over which the red sequence is assembled is
found to be a function of environment. \citet{tanaka05} follow the
build-up of the red sequence as a function of environment at three
redshifts, \mbox{$z\sim0$} (based on SDSS data), \mbox{$z=0.55$} and
\mbox{$z=0.83$}. 
In
cluster environments the data are consistent with there being no
significant growth of the red sequence since \mbox{$z=0.83$}, except for a
hint of evolution in the lowest mass bin
\mbox{($\mathcal{M}<10^{10.5}{\rm M}_{\sun}$)} between \mbox{$z=0.83$} and
\mbox{$z=0.55$}. Similarly, \citet{delucia} find evidence for a deficit of
faint \mbox{($0.1\!\la\!L^{*}\!\la 0.4$)} red sequence galaxies in 18 EDiSCS
clusters at \mbox{$0.4<z<0.8$} with respect to present day clusters. They
find that the observed decrease in the numbers of faint red sequence
galaxies are consistent with a simple model whereby these galaxies are
formed as the result of blue galaxies in clusters at \mbox{$z=0.8$}
having their star-formation suppressed by the hostile cluster environment.  
In field environments the fraction of galaxies on the red sequence is lower than in clusters at all luminosities and all
redshifts to \mbox{$z=0.8$} \citep{tanaka05}, showing that the colour-density
relation is already in place at \mbox{$z=0.8$}. There is also a deficit of
faint red sequence galaxies in field regions, both at \mbox{$z\sim0.8$} and
at the present day, indicating that the assembly of the red sequence
is still incomplete in low density environments \citep{tanaka05}.  

In \citet[Papers I and II]{paper1,paper2} we presented an analysis of
star-formation in galaxies as a function of both luminosity and environment in order to constrain the
physical mechanisms that drive the star-formation histories of
galaxies of different masses. In Paper I we used the fourth data
release of the Sloan Digital Sky Survey \citep[SDSS DR4;][]{sdssdr4} to
investigate the possible mass dependency of the SF-density and
age-density relations in the vicinity of the \mbox{$z=0.03$}
supercluster centred on the rich cluster Abell 2199. For giant
galaxies \mbox{(M$_{r}\!<\!-20$)} we found gradual age-density and SF-density
trends extending to the lowest densities studied, with the clusters
dominated by old, passively-evolving galaxies while in field regions
we found equal fractions of old, passively-evolving
and young, star-forming galaxy populations which were completely
interspersed. In contrast for the dwarf galaxy population
\mbox{($-19\!<\!{\rm M}_{r}\!<\!-17.8$)} we found a
sharp transition from the virialized regions of clusters and groups
which were still dominated by old, passively-evolving galaxies, to
outside where virtually all dwarf galaxies were young with ongoing
star-formation. The few old, passively-evolving dwarf galaxies outside
of the clusters were always found to reside in poor groups or as a
satellite to a massive galaxy. 

In Paper II we extended the analysis to the entire SDSS DR4, using a
sample of 27\,753 galaxies in the redshift range \mbox{$0.005<z<0.037$} that
is \mbox{$\ga\!9$0}\% complete to \mbox{M$_{r}=-18$}. In high-density regions
we found 70\% of galaxies to be passively-evolving
\mbox{(EW[H$\alpha]\!<\!2$\AA)} independent of luminosity. In the rarefied
field however, the fraction of passively-evolving galaxies was found
to be strongly luminosity-dependent, dropping from 50\% for
\mbox{M$_{r}\!\la\!-21$} to zero by \mbox{M$_{r}\!\sim\!-18$}. Indeed
for the lowest luminosity bin studied \mbox{($-18\!<\!{\rm
    M}_{r}\!<\!-16$)} none of the \mbox{$\sim\!6$00} galaxies in the lowest
density quartile were found to be passive. Throughout the SDSS DR4
dataset we found no passively-evolving dwarf galaxy more than two
virial radii from a massive halo, whether that be a cluster, group or
massive galaxy. 

These results imply fundamental differences in the
formation and evolution of giant and dwarf galaxies. Recently a
theoretical framework has been proposed whereby the differences are
related to: (i) the increasing star-formation efficiencies and decreasing gas
consumption time-scales with galaxy mass resultant from the
Kennicutt-Schmidt law \citep{kennicutt,chiosi}; (ii) the way that gas from the galaxy
halo cools and flows onto the galaxy \citep{keres,dekel} and which
affects its ability to maintain star-formation over many Gyr; as well
as (iii) AGN feedback which can effectively permanently shut down
star-formation in massive galaxies \citep{springel,croton}. The lack of
passively-evolving dwarf galaxies in isolated field regions implies
that {\em internal} processes, such as AGN feedback, merging or gas consumption
through star-formation, are not able to completely shut down
star-formation in these galaxies. Instead star-formation in dwarf
galaxies is only terminated once they become satellites in
massive halos, probably through the combined effects of tidal forces
and ram-pressure stripping. 

In this article we examine what consequences these differences have on the
form of the red sequence in the local universe, and its variation
from within the high-density environments of clusters and groups, to the
lowest density regions of the rarefied field. 
In $\S$~\ref{data} we describe the dataset
used, that is the same spectroscopic sample as in Paper II, which
allows us to classify the galaxies as passive, star-forming or AGN, as
well as fully define the local environment of each galaxy. In
$\S$~\ref{redseq} we examine the extent and make-up of the red
sequence as a function of environment, based on the \mbox{$u-r\,/\,{\rm M}_{r}$}
colour-magnitude diagram. In particular we examine the issue of the
contamination of the red sequence by dusty star-forming galaxies as
found by \citet{wolf} and \citet{davoodi}, and which are likely to
represent a major problem for high-redshift samples. To resolve these
problems in $\S$~\ref{galex} we take advantage of the significant coverage of the SDSS DR4
footprint by near-ultraviolet (NUV) imaging from the {\em GALEX}
public archive, to obtain \mbox{${\rm NUV}-r\,/\,{\rm M}_{r}$}
colour-magnitude diagrams in both cluster and field environments. We
show that this provides a much cleaner separation of passive and
star-forming galaxies, and find that in isolated field regions the red
sequence appears truncated at \mbox{M$_{r}\!\simeq-18.5$}, confirming the
lack of passively-evolving dwarf galaxies in these rarefied environments.
We also examine the issue of aperture bias on our classification of
galaxies, as our spectroscopic sample covers the redshift range for
which aperture biases are known to be important, at
least for massive galaxies \citep{brinchmann,kewley05}. Finally in
$\S$~\ref{discussion} we present our discussion and conclusions.
Throughout we assume a concordance $\Lambda$CDM cosmology
with \mbox{$\Omega_{M}=0.3$}, \mbox{$\Omega_{\Lambda}=0.7$} and
\mbox{H$_{0}=70\,$km\,s$^{-1}$Mpc$^{-1}$}. 

\section{The Data}
\label{data}

The galaxy sample is the same as that used and described in Paper
II, consisting of 27\,753 galaxies in the redshift range
\mbox{$0.005<z<0.037$} from the low-redshift subsample \citep{blanton05a} from the New York
University - Value Added Catalogue \citep[NYU-VAGC;][]{nyuvagc}. This is a subsample of
\mbox{$0.0033<z<0.05$} galaxies from the SDSS DR4 spectroscopic dataset for
which additional checks have been made for badly deblended objects and absolute
magnitudes have been calculated using version 3.2 of the software {\sc
  k-correct} \citep{kcorrect}. We consider only those galaxies
belonging to the two large contiguous regions in the North Galactic
Cap, excluding the three narrow stripes with
\mbox{$-60^{\circ}<\alpha<60^{\circ}$}.

We use the stellar indices of the MPA/JHU SDSS DR4 catalogues
\citep{k03a}, in which the emission-line fluxes are corrected for
stellar absorption using a continuum fitting code based on the
\citet{bc03} population synthesis models \citep{tremonti}. In Paper II
we showed that the distribution of the H$\alpha$ equivalent widths of 
\mbox{M$_{r}\!<-18$}
galaxies is strongly bimodal, allowing them to be robustly separated
into passively-evolving and star-forming populations about a value
\mbox{EW(H$\alpha)=2$\AA}. Additionally we identify AGN using the \mbox{[N{\sc
    ii}]$\lambda6584\, /\, {\rm H}\alpha$} versus \mbox{[O{\sc iii}]$\lambda 5007
\,/\, {\rm H}\beta$} diagnostic diagrams of \citet*{baldwin} as those
galaxies lying above the $1\sigma$ lower limit of the models defined
by \citet{kewley}. When either the [O{\sc iii}] or H$\beta$ lines are
unavailable (${\rm S/N}<3$), the two-line method of \citet{miller} is
used, with AGN identified as having \mbox{log([N{\sc ii}]$\lambda 6584 \,/\,
{\rm H}\alpha)>-0.2$}.

\begin{figure*}
\centerline{\includegraphics[width=17cm]{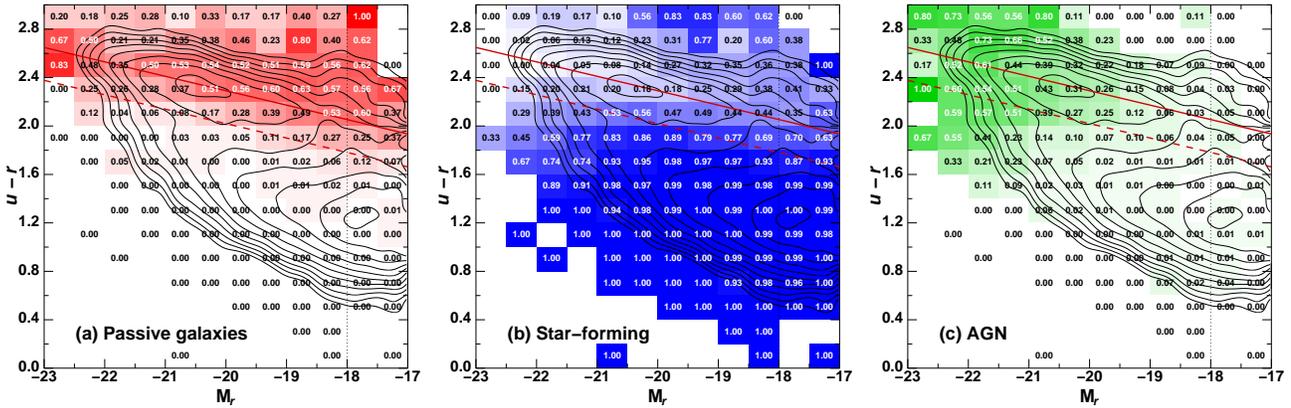}}
\caption{The global make-up of the \mbox{$u-r\,/\,{\rm M}_r$} colour-magnitude diagram of
  galaxies. Contours show the volume-corrected \mbox{$u-r\,/\,{\rm M}_{r}$}
  colour-magnitude diagram, and are spaced logarithmically, the
  spacing between contours indicating a factor $\sqrt{2}$ increase in
  the galaxy density. Panel (a) shows the global fraction of
  passively-evolving galaxies in each bin of \mbox{$u-r$} colour and M$_{r}$
  magnitude, with increasingly intense red colours indicating higher
  passive galaxy fractions in a bin. Panels (b) and (c) show
  respectively the contributions of star-forming galaxies (blue-shaded
  boxes) and AGN (green-shaded boxes).
The solid red line indicates the
  best-fit colour-magnitude relation of Eq.~1, and the dashed line
  indicates the limit used to separate red sequence and blue cloud galaxies.
}
\label{u_r_makeup}
\end{figure*}

The local environment of each galaxy is quantified using a variant of
the adaptive kernel estimator \citep{pisani96}, whereby each
\mbox{M$_{r}\!<-18$} galaxy is represented by a Gaussian kernel in
redshift-space, $K(\textbf{x},z)$, of width 500\,km\,s$^{-1}$ in the
radial direction, and whose transverse width is defined by the
distance to its third nearest neighbour within 500\,km\,s$^{-1}$. 
The
choice both of the method and of the kernel dimensions is designed
to resolve the galaxy's environment on the scale of its host dark
matter halo, as it is the mass of a
galaxy's host halo and whether the galaxy is the central or a 
satellite galaxy, that is believed to be the dominant factor in 
defining its global properties such as its star-formation history or 
morphology \citep[e.g.][]{lemson,kauffmann04,yang,blanton06}. 
In the case of galaxies within groups or clusters, the local
environment is measured on the scale of their host halo \mbox{(0.1--1\,Mpc)},
while for galaxies in field regions the local density is estimated by
smoothing over its 5--10 nearest neighbours or scales of 1--5\,Mpc. 
In Paper II we tested the efficiency of this
density estimator by applying it to the public galaxy catalogues from
the Millennium simulations \citep{springelsim}, finding it very sensitive
to the presence of even very poor groups comparable to the Local Group
containing as few as four \mbox{M$_{r}\!<-18$} 
galaxies, which represent the preferred mass-scale of major mergers
for galaxies of stellar mass \mbox{$\sim10^{10}$--1$0^{11}{\rm M}_{\odot}$} \citep{hopkins07a}.
By selecting galaxies with
\mbox{$\rho<0.5$\,Mpc$^{-2}$\,(500\,km\,s$^{-1})^{-1}$} a pure field sample
is produced with no contamination from group members. In contrast 60\%
of \mbox{$\rho>1$} and 90\%
of \mbox{$\rho>4$} galaxies lie within the virial radius of a galaxy group or cluster.

\section{The make-up of the red sequence in optical surveys}
\label{redseq}

The high quality photometric and spectroscopic data of the SDSS have
allowed thorough analyses of the bivariate colour-magnitude
distribution of galaxies to be determined, demonstrating its bimodal
nature, particularly in terms of galaxy colours
\citep[e.g.][]{strateva,baldry04}. \citet{baldry04} show that the \mbox{$u-r$}
colour distribution of galaxies in any given magnitude bin is well
described by double Gaussians for the entire magnitude range covered
\mbox{($-23.5<{\rm M}_{r}\!<-15.5$)}, allowing galaxies to be split into red
and blue populations. The relative contribution of the red sequence
population is found to increase with luminosity/mass, as well as
increasing local density \citep{baldry04,balogh04}. The relative
simplicity of observing this bimodal colour-magnitude distribution of
galaxies has allowed it to be followed out to \mbox{$z\!\sim\!1$}
\citep{bell,bundy,willmer}. To relate these observational results to
theoretical predictions for the growth and evolution of galaxies it is
broadly assumed that red sequence galaxies are passively-evolving
galaxies dominated by old stellar populations formed at \mbox{$z>1$}, while
blue galaxies are star-forming galaxies whose optical emission is
dominated by young stellar populations. The main uncertainty in such a
model is that the optical colours of galaxies do not necessarily fully
relate to the underlying star-formation history as described above. In
particular galaxies can appear red both because they are passive, but
also through high levels of dust extinction produced by star-bursts. 
A much more reliable measure of the recent star-formation history of
galaxies than their optical colour can be made from their spectra, in
particular the level of H$\alpha$ emission.  

In Figure~\ref{u_r_makeup} we look at how the \mbox{$u-r\,/\,{\rm M}_{r}$}
colour-magnitude diagram is broken up into (a) passively-evolving,
(b) star-forming and (c) AGN components as determined from their
spectra. In each panel the black contours indicate the {\em global} bivariate density
distribution of \mbox{$0.005\!<\!z\!<\!0.037$} galaxies in colour-magnitude
space. The black contours are spaced logarithmically, such that the galaxy
density doubles every two contours.
Over this redshift range the SDSS spectroscopic sample is
expected to be \mbox{$\ga\!9$0\%} complete to \mbox{M$_{r}\!=-18.0$}, the main source of incompleteness
being due to fibre collisions, but at fainter magnitudes galaxies will
only be observable within a certain redshift range, which in some
cases may be much less than overall redshift range of the sample. To
correct for this issue, we weight each galaxy by V$_{survey}/{\rm
  V}_{max}$ where V$_{max}$ is the maximum volume over which the
galaxy could be observed within the survey volume, V$_{survey}$.  

For each bin in \mbox{$u-r$} colour and M$_{r}$ magnitude we calculate the
relative contribution of passively-evolving galaxies (\mbox{EW[H$\alpha]\!<\!2$}\AA; not
AGN), star-forming galaxies (\mbox{EW[H$\alpha]\!>\!2$}\AA; not AGN) and AGN, in
panels (a-c) respectively. For each bin in the colour-magnitude
diagram containing at least one galaxy, the {\em global}
fraction of passively-evolving galaxies is indicated both by the
numeric value, and the colour shading of the box, with increasingly
intense red colours indicating higher passive galaxy fractions in that
bin. Panels (b) and (c) show respectively the contributions of
star-forming galaxies (blue-shaded boxes) and AGN (green-shaded boxes).

The bimodality of galaxy properties in colour-magnitude space is
clear, with a population of red \mbox{($u-r\sim2.4$)} galaxies that extends
over the full magnitude range covered forming the red sequence, and a
second population of blue \mbox{($u-r\sim1.2$)} galaxies that show a greater
dispersion in colour, but tend to be less luminous in general than
their red counterparts. Panel (a) shows that passively-evolving
galaxies are well confined to the red sequence, with virtually no
passive galaxies showing blue optical colours with
\mbox{$u-r\la1.6$}.

The C-M relation of red sequence galaxies shows the well known slope
due to metallicity effects. We apply the method of \citet{lopez} to estimate the slope and width of the C-M relation, using the biweight algorithm
\citep{beers} to estimate the dispersion about the relation, and
varying the slope to minimize the biweight scale of the deviations
about the median \citep[see e.g.][]{haines04}. Considering just the passively-evolving galaxies we
obtain a best-fitting relation of:
\begin{equation}
u-r=2.291\,(0.004)-0.1191\,(0.0114) \times({\rm M}_{r}+20)
\end{equation}
(shown by the red solid lines in panels a-c), and
a width $\sigma=0.181$. The best fit closely follows the main axis of
the maximum in the galaxy density contours due to the red sequence
population, demonstrating the validity in using only the passive
galaxies to obtain the fit.
We identify red sequence galaxies as those
which have colours redder than the black dashed line, which
corresponds to a relation \mbox{$1.5\,\sigma$} bluer than the red sequence, and
where equal numbers of passive and star-forming galaxies are observed
when averaged over all environments.
    
\begin{figure*}
\centerline{\includegraphics[width=12cm]{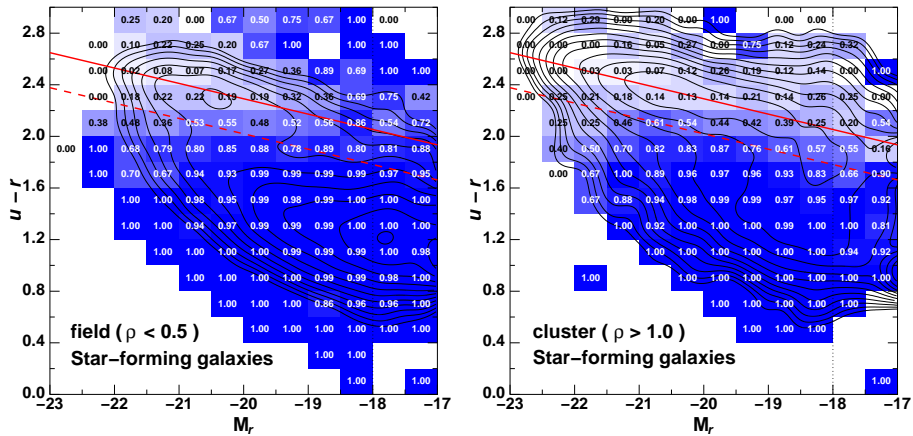}}
\caption{The make-up of the \mbox{$u-r\,/\,{\rm M}_r$} colour-magnitude diagram in
  field (\mbox{$\rho<0.5$}; left panel) and cluster (\mbox{$\rho>1$};
  right panel) environments. Contours show the volume-corrected
  \mbox{$u-r\,/\,{\rm M}_{r}$}
  colour-magnitude diagram, and are spaced logarithmically, the
  spacing between contours indicating a factor $\sqrt{2}$ increase in
  the galaxy density. Each panel shows the fraction of
  star-forming galaxies in each bin of $u-r$ colour and M$_{r}$
  magnitude, with increasingly intense blue colours indicating higher
  passive galaxy fractions in a bin. 
The solid and dashed red lines are the same as in Fig.~\ref{u_r_makeup}.
}
\label{makeup_rho}
\end{figure*}

In panel (b) we see that virtually all blue galaxies are
identified as star-forming from their spectra, and hence to a first
approximation the correlation between galaxy colour and the current
star-formation rate appears good. 
However it is also
notable that a significant fraction of galaxies with red optical
colours \mbox{($u-r\ga2$)} are in
fact star-forming, particularly at the faint end \mbox{($-19<{\rm
  M}_{r}\!<-17$)}, where 53\% of red sequence galaxies (i.e. above
the dashed line) are spectroscopically classed as star-forming. 
Finally in panel (c) we see that AGN are generally red and confined to
the most luminous galaxies in the sample where they make up \mbox{$\sim\!5$0\%}
of the galaxy population, this fraction falling to
close to zero by \mbox{M$_{r}\!\sim-18$} as shown in Paper II.

Hence, while passive galaxies are red, {\em red galaxies are not
  necessarily passive.} 
In an analysis of the SDSS main sample galaxies covered by infrared
imaging from the SWIRE survey, \citet{davoodi} find that 17\% of red
sequence galaxies are dusty, star-forming galaxies (identified by
their high 24\,$\mu$m to 3.6\,$\mu$m flux ratios and H$\alpha$ emission),
while \citet{wolf} find that dusty, star-forming galaxies make up more
than one-third of the red sequence population in the the A\,901/2
supercluster system at $z=0.17$, finding them preferentially
in the medium-density outskirts of the clusters. 

Given that the relative fractions of passively-evolving/red and star-forming/blue galaxies of a given
luminosity/stellar mass are strongly dependent on local environment,
particularly at the faint end \citep{balogh04,baldry06,paper2}, the
contamination of the red sequence by star-forming galaxies is also
likely to be strongly dependent on environment. In
Fig.~\ref{makeup_rho} we show how the optical colour-magnitude diagram
and the relative contribution from star-forming galaxies changes from
field (\mbox{$\rho<0.5$}; left panel) to cluster (\mbox{$\rho>1.0$};
right panel) environments. 
The contours in each panel indicate the bivariate volume-corrected
number density in colour-magnitude space as in Fig.~\ref{u_r_makeup},
this time 
for field and cluster galaxies respectively. The bivariate distribution of field galaxies
is similar to that for the global population, mainly as the majority (64\%)
of galaxies are found in field regions. The main difference is that in
field regions the red sequence is less well populated, particularly at
faint magnitudes. In cluster regions, the bivariate colour-magnitude
distribution of galaxies is instead dominated by the red sequence,
which can be clearly seen to extend to the completeness limit of the
sample at \mbox{M$_{r}\!=-18$}, the ``blue cloud'' appearing now merely as a
tail of galaxies extending bluewards from the red sequence.

Looking now at the fraction of red sequence galaxies which are in fact
spectroscopically classed as star-forming, it is clear that the
contamination of dusty star-forming galaxies is much greater in field
regions than in clusters, and in all environments the contamination is
greater at faint magnitudes. It is also notable that most galaxies
that are above the red sequence \mbox{($u-r\sim2.8$)} are in fact
star-forming, their red colours simply due to dust extinction. 

We
summarize these trends in Table~\ref{fracs} which shows the fraction
of red sequence galaxies (i.e. above the red dashed lines in
Figures~\ref{u_r_makeup} and~\ref{makeup_rho}) which are spectroscopically classed as
star-forming as a function of both environment and luminosity. 
Globally, we find that 31\% of red sequence galaxies (with
\mbox{M$_{r}\!<-17$}) are star-forming, similar to that found by \citet{wolf},
but larger than the 17\% obtained by \citet{davoodi}. This difference
is most likely due to their sample being biased towards more luminous
galaxies than ours, and in fact we find just 18\% of \mbox{M$_{r}\!<-20$} galaxies to be
star-forming. 

\begin{table}
\begin{center}
\begin{tabular}{cccc} \\ \hline
Magnitude & Global & Field & Cluster\\
range & fraction & ($\rho<0.5$) & ($\rho>1.0$) \\ \hline
all & 31.$4\pm1.1$\% & 3$7.2\pm2.2$\% & 24.$6\pm1.5$\% \\
M$_{r}\!<-20$ & 1$8.1\pm1$.1\% & 21.$0\pm2.1$\% & 14.$0\pm0.4$\% \\
\mbox{$\!-19<{\rm M}_{r}\!<-17\!$} & 5$1.7\pm3.1$\% & 7$3.0\pm7.8$\% & 3$5.4\pm3.5$\% \\ \hline
\end{tabular}
\end{center}
\caption{Fraction of red sequence galaxies classified as star-forming
  as a function of both environment and luminosity}
\label{fracs}
\end{table}

The most notable figure in Table~\ref{fracs} is that for
\mbox{$-19<{\rm M}_{r}\!<-17$} red sequence galaxies in field regions, where we
find 7$3\pm8$\% to be classed as star-forming. This has important
consequences for interpreting many of the recent studies for the
build-up of the red sequence, both in the local universe, and at high
redshifts. For example, \citet{baldry06} show that in low-density
regions the fraction of galaxies belonging to the red sequence is
strongly dependent on stellar mass, dropping from \mbox{$\sim\!8$0\%} at \mbox{$\log
\mathcal{M}\!\sim\!11.4$} to \mbox{$\sim\!5$\%} at
\mbox{$\log\mathcal{M}\!\sim\!9.0$}. However in a similar analysis
where galaxies are spectroscopically classified as passive or star-forming
from their H$\alpha$ emission, \citet{paper2} find that the fraction
of passively-evolving galaxies drops to precisely zero by
\mbox{M$_{r}\!\sim-18$} or \mbox{$\log\mathcal{M}\!\sim\!9.2$}.
Similarly, \citet{tanaka05} determine the luminosity function of red
sequence galaxies in field regions, obtaining a Schechter function
with a very shallow faint-end slope ($\alpha=-0.14$), but which
appears to flatten out at the faintest magnitudes \mbox{($-19<{\rm
  M}_{V}\!<-17.5$)} due to a residual population of faint red
sequence galaxies. Our results indicate that around three-quarters of
this residual faint red sequence population in field regions are in
fact star-forming galaxies. Hence if a more robust colour selection
could be applied to separate passive and star-forming galaxies, this
residual faint red population would largely disappear in field regions
(if not entirely\footnote{In Paper II, we find all the few passively-evolving
dwarf galaxies in field regions to be satellites to massive
galaxies}), and the resultant luminosity function of red sequence
galaxies would have an even shallower faint-end slope, and could even
be completely truncated.

\begin{figure*}
\centerline{\includegraphics[width=17cm]{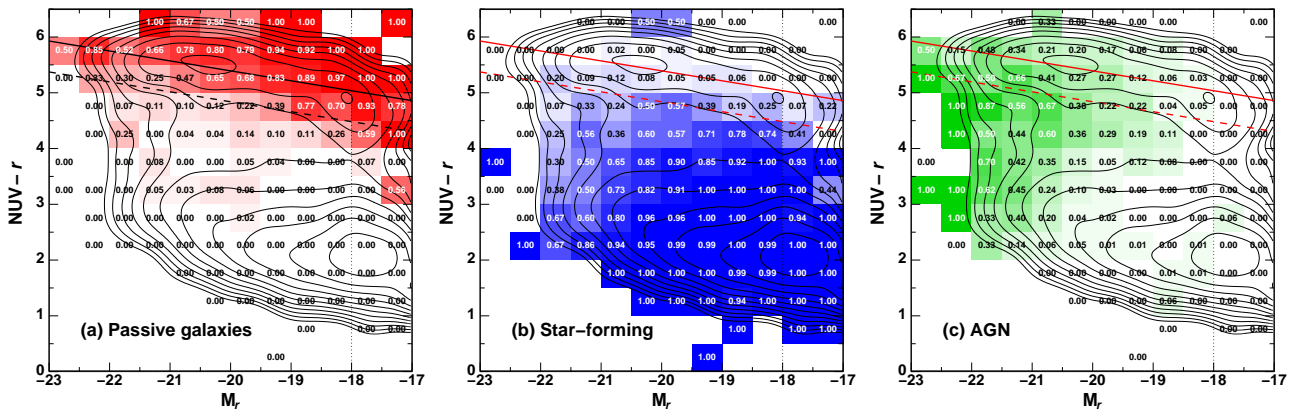}}
\caption{The global make-up of the \mbox{${\rm NUV}-r\,/\,{\rm M}_r$} colour-magnitude diagram of
  galaxies. Contours show the volume-corrected \mbox{${\rm NUV}-r\,/\,{\rm M}_{r}$}
  colour-magnitude diagram, and are spaced logarithmically, the
  spacing between contours indicating a factor $\sqrt{2}$ increase in
  the galaxy density. Panel (a) shows the global fraction of
  passively-evolving galaxies in each bin of \mbox{${\rm NUV}-r$} colour and M$_{r}$
  magnitude, with increasingly intense red colours indicating higher
  passive galaxy fractions in a bin. Panels (b) and (c) show
  respectively the contributions of star-forming galaxies (blue-shaded
  boxes) and AGN (green-shaded boxes).
  The solid red line indicates the
  best-fit colour-magnitude relation of Eq.~2, and the dashed line
  indicates the limit used to separate red sequence and blue cloud galaxies.
}
\label{NUV_r_makeup}
\end{figure*}

\section{The GALEX-SDSS view of the red sequence at $\bmath{z\sim0}$}
\label{galex}

To resolve the limitations inherent in the SDSS $u-r$ colour-magnitude
diagram in separating passively-evolving and star-forming galaxies, we
consider instead the \mbox{${\rm NUV}-r\,/\,{\rm M}_{r}$} colour-magnitude
relation. The launch of the {\em Galaxy Evolution Explorer}
\citep{martin05} has
allowed both near-ultraviolet  (NUV; \mbox{$\lambda_{\rm eff}=2316$\AA},
\mbox{$\Delta\lambda=732$\AA}) and far-ultraviolet (FUV;
\mbox{$\lambda_{\rm eff}=1539$\AA}, \mbox{$\Delta\lambda=268$\AA}) photometry to be
obtained for a large sample of galaxies from the SDSS. The UV
photometry from {\em GALEX} provides a global measure of
star-formation on timescales \mbox{$\sim\!10^{8}$} years, that is an order of
magnitude more sensitive to low levels of recent star-formation
(where \mbox{$\la\!1$\%} of the mass in stars in the galaxy form in the last Gyr)
overlaid on otherwise old stellar populations than optical photometry
 (e.g. $u-r$) alone
\citep[e.g.][]{martin05,martin07,kauffmann06}. Similarly,
the \mbox{${\rm NUV}-r$} colour is shown to correlate tightly
with the birthrate parameter $b$ \citep[the ratio of the current to
past-averaged star-formation rates;][]{salim05} and the age-sensitive
spectral indices d4000 and H$\delta$ \citep{martin07},   
implying that the star-formation
history of a galaxy can be constrained from its \mbox{${\rm NUV}-r$} colour
alone.  
Indeed analyses of the
UV-optical colour-magnitude relations of early-type galaxies selected
from the matched GALEX/SDSS catalogues found a much greater scatter in
the UV-optical colours than from their optical colours, a result that
was interpreted as due to low-level residual star-formation being
common in early-type galaxies
\citep{rich,kaviraj,schawinski}. Moreover \citet{salim} show that the
combined FUV and NUV photometry can be used to obtain reliable
measures of dust-corrected SFRs for star-forming galaxies to an
accuracy of 0.2\,dex that agree with estimates based on H$\alpha$
fluxes to 10\% across the entire range of galaxy masses.
   
For this analysis we 
use the 3rd data
release of the {\em Galaxy Evolution Explorer} (GALEX GR3)
near-ultraviolet 
imaging \citep{martin05}. Passively-evolving galaxies at \mbox{$z\sim0$} have
\mbox{${\rm NUV}-r\sim5$.5--6} \citep{yi}, requiring \mbox{$m_{\rm NUV}\!\ga23$} imaging to detect all galaxies
from our 
\mbox{$0.005<z<0.037$} SDSS spectroscopic sample. Hence we consider those
GALEX GR3 images from the Medium Imaging Survey (MIS), Nearby
Galaxy Survey (NGS) and the publicly available Guest Investigator
images which have exposure times $\sim1\,$ksec and
$5\,\sigma$ depths of \mbox{$m_{\rm NUV}\!\sim23$}. We also use images
from the Deep Imaging Survey (DIS) and two fields from Guest Investigator
Program 35
(PI: G. Williger) which have exposure times \mbox{$\sim\!30$\,ksec} and $5\,\sigma$ depths of \mbox{$m_{\rm NUV}\!\sim24.5$}
\citep{martin05}. In total 528 GALEX GR3 NUV images from these surveys overlap with the SDSS DR4 footprint, for a total area of 490.1\,deg$^{2}$.

For each galaxy in our \mbox{$0.005<z<0.037$} catalogue that is covered by
deep GALEX NUV imaging, we take the {\em best} (using the best
available reduction which may not be that used to select the
spectroscopic target) SDSS $r$-band ``corrected
frame'' \mbox{$2048\times1489\,{\rm pix}^{2}$}
\mbox{($13.5^{\prime}\times9.8^{\prime}$)} image from the SDSS Data Archive
Server (DAS). If there is more than one image, we take the one where the galaxy is furthest from the
image boundaries. We then register the associated GALEX NUV image with
the SDSS $r$-band image using the {\sc iraf} tool {\sc
  sregister}. This takes as input the pipeline astrometric
calibrations for each image, the relative astrometric precision
between the two images being 0.49\,arcsec \citep{morrissey07}.
We obtain the integrated galaxy \mbox{${\rm NUV}-r$} colour, by running SExtractor in
dual-image mode, determining the colour over the Kron aperture 
determined from the $r$ image. 
This step is necessary as 
the light distribution from galaxies can be quite different between
the NUV and $r$-bands: the dominant source of NUV flux are
star-forming regions in the disks, while that the $r$-band flux comes
more from the old stellar populations in the bulge. In late-type
spirals in particular, the NUV flux distribution may be shredded by
SExtractor into separate star-forming regions. Hence simply matching
the nearest GALEX NUV detection with the SDSS $r$-band detection can
result in artificially red colours as the NUV aperture does not cover
the whole galaxy. 
We use
the SDSS and GALEX photometric calibrations
\citep{stoughton,morrissey} which have zero-point uncertainties of
0.01 and 
0.03\,m$_{\rm AB}$ in the $r$ and NUV bands
respectively \citep{morrissey07}. We make no correction for the
different point-spread functions (PSFs) of the NUV and $r$-band images, the
typical FWHMs being 4.9 and 1.0--1.5 arcsec respectively. This can
result in a loss of flux due to the extended PSF in the NUV-band, but
for the typical-sized Kron apertures in our sample the effects should be
less than 0.10\,mag \citep{morrissey07}.  
We correct for Galactic
extinction using A(NUV$)=8.18\,{\rm E}(B-V)$ and A($r)=2.751\,{\rm
  E}(B-V)$ using the dust extinction maps of \citet{schlegel}. 

In total 4065 galaxies from our \mbox{$0.005\!<\!z\!<\!0.037$} sample were covered by
the GALEX NUV imaging, of which just two were not detected, both of
which were passively-evolving dwarf galaxies \mbox{(M$_{r}\!\sim-18$)} in
high-density regions \mbox{($\rho>1$)}. 

In Figure~\ref{NUV_r_makeup} we show the global make-up of the \mbox{${\rm
  NUV}-r\,/\,{\rm M}_{r}$} colour-magnitude diagram, breaking it up into
  its (a) passive-evolving, (b) star-forming and (c) AGN components,
  analogously to the optical colour-magnitude diagram of
  Fig.~\ref{u_r_makeup}. As before, the contours indicate the global
  volume-weighted bivariate number distribution of \mbox{$0.005<z<0.037$}
  galaxies. The bimodality of galaxy properties in colour-magnitude
  space is again apparent, with a robust separation of red and blue
  galaxies about a colour \mbox{${\rm NUV}-r=4$} that extends over the full magnitude
  range covered \citep[see also][]{wyder}, and appears much cleaner than for the optical
  counterpart. The separation of the red sequence and the ``blue
  cloud'' is \mbox{$\sim\!3$\,mag} in \mbox{${\rm NUV}-r$}, but only
  \mbox{$\sim\!0.9$\,mag} in \mbox{$u-r$}. 

Panel (a) shows that, as for the
  optical colour-magnitude diagram,
  passive galaxies are well confined to the red sequence, with few
  passive galaxies showing blue UV-optical colours with \mbox{${\rm
  NUV}-r\la4$}. We estimate the zero-point, slope and dispersion of the red sequence
as before using the ${\rm NUV}-r$ colours and $r$-band absolute
magnitudes of each galaxy classified as passively-evolving
  \mbox{(EW[H$\alpha]<2$\AA}; including AGN), obtaining
a C-M relation of:
\begin{equation}
{\rm NUV}-r =  5.393\,(0.013) - 0.1782\,(0.0223) \times ({\rm M}_{r}+20.)
\end{equation}
(shown by the black solid line in panel a)
and a dispersion of \mbox{$\sigma_{{\rm NUV}-r}=0.370\pm0.007$}. This
value is consistent with the 0.3--0.5 mag obtained by \citet{wyder}
who fitted the \mbox{${\rm NUV}-r$} colour distribution for each half
magnitude bin in M$_{r,0.1}$ by two Gaussian functions for the red and
blue sequences. Thus the scatter in the NUV-optical red
sequence of \citet{wyder} is principally due to the scatter from the
passively-evolving component which dominates it. 
Our $\sigma$ is notably smaller than the \mbox{$\sim1$\,mag} dispersion
observed by \citet{kaviraj} for morphologically selected early-type galaxies
(with the SDSS parameter $frac\_Dev > 0.95$ in $g$, $r$ and $i$ bands)
from the SDSS DR3, due to a significant fraction of these
early-type galaxies having ongoing star-formation.
As before we
identify red sequence galaxies as those which are less than
\mbox{$1.5\,\sigma$} bluer than the C-M relation, i.e. that lie above the dashed line.

\begin{figure*}
\centerline{\includegraphics[width=17cm]{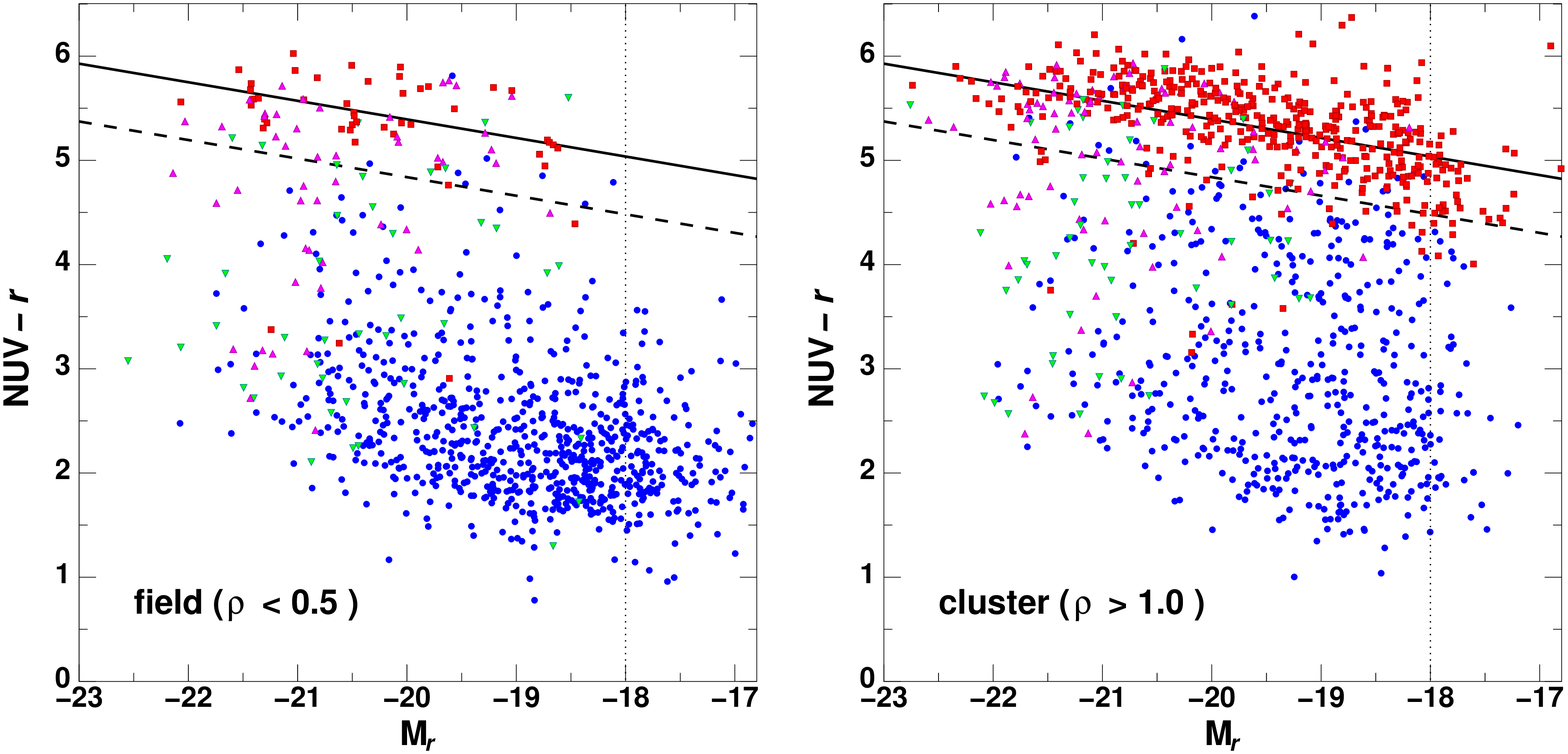}}
\caption{The \mbox{${\rm NUV}\,/\,{\rm M}_r$} colour-magnitude diagram of galaxies in the
  lowest \mbox{($\rho<0.5$)} and highest \mbox{($\rho>1.0$)} density
  environments. Red squares and blue circles indicate passive
  \mbox{(EW[H$\alpha]<2$\AA)} and star-forming \mbox{(EW[H$\alpha]>2$\AA)}
  galaxies, while green and magenta triangles indicate AGN with
  \mbox{EW(H$\alpha)>2$\AA}\, and \mbox{EW(H$\alpha)<2$\AA}\, respectively. The dotted line indicates the
  completeness limit of the volume covered, while the solid and dashed
  lines are as in Fig.~\ref{NUV_r_makeup}.}
\label{galexcm}
\end{figure*}

In panel (b) we see that virtually all blue galaxies with \mbox{${\rm
  NUV}-r<4$} are spectroscopically star-forming (except at the very
  bright end where AGN begin to dominate as shown in panel c). Unlike the optical
  colour-magnitude diagram where a significant fraction of red
  galaxies were also classified as star-forming, we see few
  star-forming galaxies with \mbox{${\rm NUV}-r\ga5$}. Indeed, globally
  we find that just 8\% of galaxies belonging to the NUV-optical red
  sequence are classified as star-forming as opposed to 31\% of
  galaxies from the optical red sequence. In Table~\ref{galex_fracs}
  we report the corresponding fractions of star-forming galaxies among
  the NUV-optical red sequence population as a function of both
  magnitude and environment. In all cases the contamination of the
  NUV-optical red sequence by star-forming galaxies is significantly
  less (by typically a factor 3--5) than for the optically selected red sequence galaxies, being 
  \mbox{$\la\!1$0\%} for all subsamples except for the faint, field galaxy
  population where 30\% appear to be star-forming. However we note
  that this last result is based on just four star-forming galaxies
  out of thirteen \mbox{$-19\!<\!{\rm M}_{r}\!<\!-17$} red sequence galaxies.  
We thus argue that the position of a galaxy in the \mbox{${\rm NUV}-r\,/\,{\rm
  M}_{r}$} colour-magnitude diagram can be used as a robust classifier
  of its recent star-formation history, efficiently separating
  passively-evolving and star-forming galaxies. 

In panel (c) we show that AGN dominate the bright end of the
colour-magnitude diagram, and typically have \mbox{${\rm NUV}-r$} colours
that are intermediate between the red sequence and blue cloud, lying
in the so-called ``green valley''. Indeed this preponderance of AGN in
the transition zone between the red sequence and blue cloud has been
proposed as evidence for galaxies becoming passive as the result of
AGN feedback \citep[e.g.][]{kauffmann06,martin07}. The concept that
these intermediate objects in the ``green valley'' represent a
third population that is distinct both from the red and blue sequences
is confirmed by \citet{wyder} who were unable to fit the overall
\mbox{${\rm NUV}-r$} colour distribution of galaxies by double
Gaussians in the same way as \citet{baldry04} were able to using their $u-r$
colours, due to excess galaxies with intermediate \mbox{${\rm NUV}-r$} colours.
Moreover, they show that this ``green valley'' population remains even
after correcting their colours for dust exctinction, confirming that
they represent a true intermediate population.

\begin{table}
\begin{center}
\begin{tabular}{cccc} \\ \hline
Magnitude & Global & Field & Cluster\\
range & fraction & \mbox{($\rho<0.5$)} & \mbox{($\rho>1.0$)} \\ \hline
all & 8.$5\pm1.3$\% & 10.$6\pm4.2$\% & 5$.8\pm1.$3\% \\
M$_{r}<-20$ & 6.$1\pm1.$6\% & 3$.3\pm2.7$\% & 5.$0\pm1.9$\% \\
\mbox{$\!-19<{\rm M}_{r}\!<-17\!$} & 10.$2\pm2$.7\% & 30.$8\pm23.9$\% & 5$.0\pm2.$0\% \\ \hline
\end{tabular}
\end{center}
\caption{Fraction of \mbox{${\rm NUV}-r$} selected red sequence galaxies classified as star-forming as a function of both environment and luminosity}
\label{galex_fracs}
\end{table}

In Paper II we found that the fraction of passively-evolving
galaxies is a strong function of both luminosity/stellar mass and
local environment. In particular we found that in low-density regions
corresponding to field environments well outside the influence of
clusters and groups, the fraction of passively-evolving galaxies is a
strong function of luminosity, dropping from \mbox{$\sim\!5$0\%} for
\mbox{$\sim\!{\rm M}^{*}$} galaxies to zero by
\mbox{M$_{r}\!\sim-18$}. Moreover, those few passively-evolving galaxies
outside groups and clusters were invariably found to be satellites to
massive galaxies. This was put forward as strong evidence that dwarf
galaxies do not become passive through internal mechanisms, but only
through environment-related processes after they become satellites
within massive halos. 

In Figure~\ref{galexcm} we compare the \mbox{${\rm NUV}-r\, /\, {\rm M}_{r}$}
colour-magnitude diagrams of galaxies in field (\mbox{$\rho<0.5$}; left panel) and
cluster (\mbox{$\rho>1.0$}; right panel) environments. In Paper II we found
that the few passively-evolving dwarf galaxies in field regions
\mbox{($\rho<0.5$)} were almost always satellites to massive galaxies
\mbox{(M$_{r}\!<-20$)}, and that throughout the SDSS DR4 dataset there were no
passively-evolving dwarf galaxies more than two virial radii from a
massive halo, whether that be a cluster, group or massive galaxy.
Hence to confirm this result we remove those field galaxies within a
projected distance of 400\,kpc and a radial velocity within
300\,km\,s$^{-1}$ of a \mbox{M$_{r}\!<-20$} galaxy. These comprise
around 10\% of the total field galaxy population.
Our resultant field galaxy sample thus only contains galaxies that are unlikely
to have had any encounter with a massive halo in their past. The
NUV-optical red sequence should then represent only those galaxies which
have become passive through internal processes such as merging, AGN
feedback and gas exhaustion through star-formation.

We separate
galaxies into 
star-forming (\mbox{EW[H$\alpha]>2$\AA;} shown as blue circles) and
passively-evolving (\mbox{EW[H$\alpha]<2$\AA;} red squares) galaxies, to
measure the correspondence between the current star-formation rate and the
\mbox{${\rm NUV}-r$} colour of galaxies. Galaxies with optical AGN
signatures are indicated by green triangles (for those with
\mbox{EW[H$\alpha]>2$\AA)} or magenta triangles \mbox{(EW[H$\alpha]<2$\AA)}.

In cluster regions (\mbox{$\rho>1$}; right panel) there is a clear NUV-optical red
sequence dominated by passively-evolving galaxies that extends to at
least the completeness limit of the survey \mbox{(M$_{r}\!=-18$)}. There appears to be a
break in the red sequence at \mbox{M$_{r}\!\sim-20.5$}, with massive red
sequence galaxies having similar colours \mbox{(${\rm NUV}-r\sim5.7$)} 
while fainter galaxies become increasingly blue. Similar trends have
been seen by \cite{boselli} who find the dichotomy between giant and
dwarf red sequence galaxies to be even stronger when using the
FUV-optical or FUV-NIR colours. \citet{wyder} in contrast find no
evidence of a break in the NUV-optical red sequence, fitting it instead by a
straight line for \mbox{$-23.5<{\rm M}_{r,0.1}<-18$}. 
 Fitting separate C-M relations to
bright \mbox{(M$_{r}<-20$)} and faint \mbox{($-21<{\rm M}_{r}\!<-18$)}
passively-evolving galaxies we obtain slopes of $-0.043\pm0.052$ and
$-0.222\pm0.039$ respectively, and hence the break appears significant. 
Considering now a C-M relation with break at \mbox{M$_{r}\!=-20.5$}
  the overall dispersion is reduced slightly to \mbox{$\sigma_{{\rm
      NUV}-r}=0.345\pm0.007$}, with no significant difference between 
  bright and faint magnitudes.  

At all luminosities we see an extended
tail of galaxies with bluer colours, being mostly AGN at the
bright-end and star-forming galaxies at fainter magnitudes. Although
many of these galaxies have \mbox{${\rm NUV}-r\sim2$} corresponding to the
peak of the blue cloud distribution seen in Fig.~\ref{NUV_r_makeup}, a
significant fraction are found with transitional colours being in the
``green valley'' with \mbox{${\rm NUV}-r\sim4$--5}. We see no
obvious separation between the red sequence and the blue cloud
populations in cluster environments.

In isolated field regions however, the dominant feature of the
colour-magnitude diagram becomes the blue cloud population at
\mbox{${\rm NUV}-r\sim2$}, made up almost entirely of star-forming
galaxies, except at the very bright tip \mbox{(M$_{r}\!\la{\rm
    M}^{*}$)} where AGN take over. The red sequence is now sparsely
populated at all magnitudes and appears truncated below \mbox{M$_{r}\!\sim-18.5$}. 

\subsection{The red and blue galaxy luminosity functions}

The galaxy luminosity function, which describes the number of galaxies
per unit volume as a function of luminosity, is a powerful tool to constrain galaxy evolution and transformations, since it is directly related to the
galaxy mass function. In particular, by separating galaxies by their
colour into the red sequence and blue cloud populations, and obtaining
the type-specific luminosity functions, the growth of the red sequence
and the transformation of blue to red galaxies can be followed.  

\begin{figure}
\centerline{\includegraphics[width=8cm]{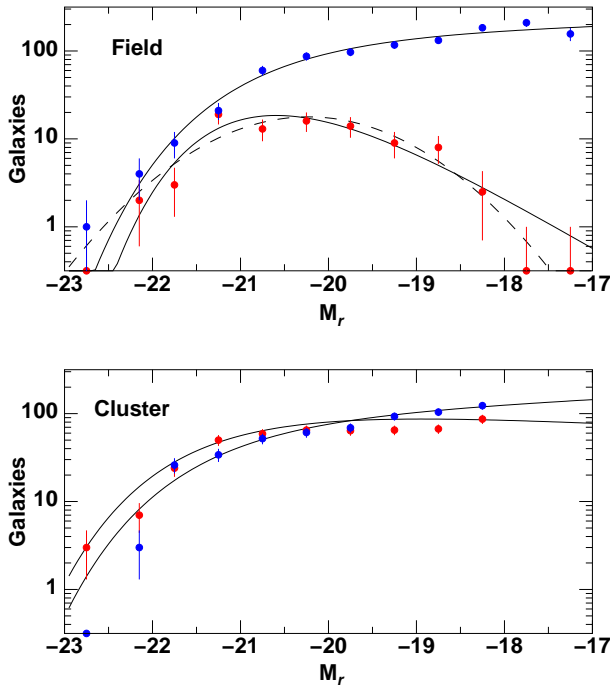}}
\caption{The red and blue galaxy luminosity functions for cluster and
  field environments. The best-fit single Schechter functions are
  indicated by black curves. For the red sequence galaxies in field
  regions the best fitting single Gaussian function is also shown by
  the dashed black curve.} 
\label{lfs}
\end{figure}

\begin{figure}
\centerline{\includegraphics[width=8cm]{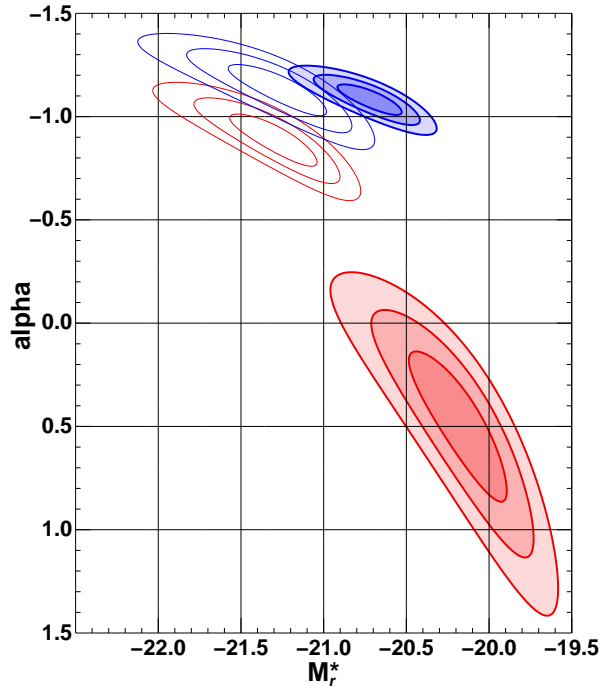}}
\caption{The confidence limits for the best-fitting Schechter functions
to each of the four luminosity functions shown in Fig.~\ref{lfs}. The
red contours indicate the 1, 2 and \mbox{$3\,\sigma$} confidence limits for
the parameters $\alpha$ and M$_{r}$ for the ${\rm NUV}-r$-selected red
sequence LF, while the blue contours indicate the corresponding
confidence limits for the blue cloud population. The filled and empty
contours correspond respectively to the galaxy luminosity functions in
field and cluster environments.
} 
\label{lfparams}
\end{figure}

In Figure~\ref{lfs} we plot the red and blue volume-corrected galaxy luminosity
functions (as red and blue symbols respectively) in both field and cluster environments, based on the
galaxies shown in Fig.~\ref{galexcm}, where the dashed line is used to
separate the red and blue galaxy populations. We correct for incompleteness at
faint magnitudes as before by weighting each galaxy by
V$_{survey}/{\rm V}_{max}$. For each luminosity function we determine
the best-fitting single \citet{schechter} function based on a
maximum-likelihood analysis, shown by the black curves.
The resultant best-fit parameters are presented in
Table~\ref{lumparams}, where the reported errors represent the 68.3\%
confidence limits in $\alpha$ and M$^{*}$. The uncertainties in
$\alpha$ and M$^{*}$ are expected to be correlated, and in
Figure~\ref{lfparams} we present the 1, 2 and 3\,$\sigma$ confidence
limits in M$_{r}^{*}$ and $\alpha$ defined as the regions containing
68.3, 95.4 and 99.7\% of the probability of finding the parameters within
the contours. The filled and empty contours correspond respectively to
the galaxy luminosity functions in field and cluster environments. 

We see little sign of variations in the luminosity function of
blue galaxies from field to cluster environments, with marginal
evidence for an increase in the luminosity of M$^{*}$ from field to
cluster environments, but no difference in the faint-end slope.
In contrast the luminosity function of the red sequence populations in
field and cluster environments are inconsistent at \mbox{$>\!3\,\sigma$}
level. In particular, the red sequence luminosity function in field
regions peaks at \mbox{M$_{r}\!\sim-20.5$} and drops rapidly at magnitudes
fainter than \mbox{M$_{r}\!\sim-18.5$}, with no red galaxies at
\mbox{M$_{r}\!>-18$}. This decline results in a very shallow
faint-end slope \mbox{$\alpha=+0.484\pm0.365$}, and we 
are also able to fit the red sequence field galaxy population
equally well by a
single Gaussian function with \mbox{$\sigma=0.684\pm0.088$} and
\mbox{M$_{r}^{*}\!=-20.23\pm0.16$} as shown by the dashed curve in Fig~\ref{lfs}. 
Although we are limited by the small number of red galaxies in our
SDSS-GALEX field
sample, it is possible that the red sequence is completely truncated
below \mbox{M$_{r}\!\ga-18$} in these rarefied environments.

\begin{table}
\begin{center}
\begin{tabular}{ccccc} \hline
LF params & $\alpha$ & M$^*$ & $\psi$ \\ \hline
\multicolumn{4}{l}{Field regions ($\rho<0.5$, $>\!40$0\,kpc from
  \mbox{M$_{r}\!<\!20.0$} galaxy)} \\ 
red & $+0.484\pm0.365$ & $-20.16\pm0.26$ & 49.3 \\
blue & $-1.078\pm0.071$ & $-20.71\pm0.19$ & 163.2 \\ \hline
\multicolumn{4}{l}{Cluster regions ($\rho>1.0$)} \\ 
red & $-0.884\pm0.122$ & $-21.28\pm0.25$ & 135.9 \\ 
blue & $-1.126\pm0.121$ & $-21.25\pm0.28$ & 110.3 \\ \hline 
\end{tabular}
\end{center}
\caption{Best-fitting luminosity functions}
\label{lumparams}
\end{table}

\subsection{Aperture biases in the SDSS spectroscopic sample}

One of the major concerns of using the fibre-obtained SDSS spectra
for nearby galaxy samples, such as those used in Papers I and II, is
the effect of aperture biases, due to the spectra being obtained
through 3 arcsec diameter apertures rather than over the full extent
of the galaxy. Significant radial star-formation gradients are
possible within galaxies, particularly those undergoing nuclear
star-bursts or spiral galaxies with prominent passively-evolving
bulges, that can result in the ``global'' star-formation rate
extrapolated from spectra containing flux from only the galaxy
nucleus being significantly over or underestimated. \citet*{kewley05}
indicate that star-formation rates based on spectra obtained through
apertures containing less than \mbox{$\sim\!2$0\%} of the integrated
  galaxy flux can be over or underestimated by a factor
  \mbox{$\sim\!2$}, and to ensure the SDSS fibres sample more than
  20\% require galaxies to be at \mbox{$z>0.04$}. Clearly in order to use the
  SDSS dataset to study star-formation in dwarf galaxies this is not
  possible, as at $z=0.04$ they are already too faint to be included
  in the SDSS spectroscopic sample.

\begin{figure}
\centerline{\includegraphics[width=8cm]{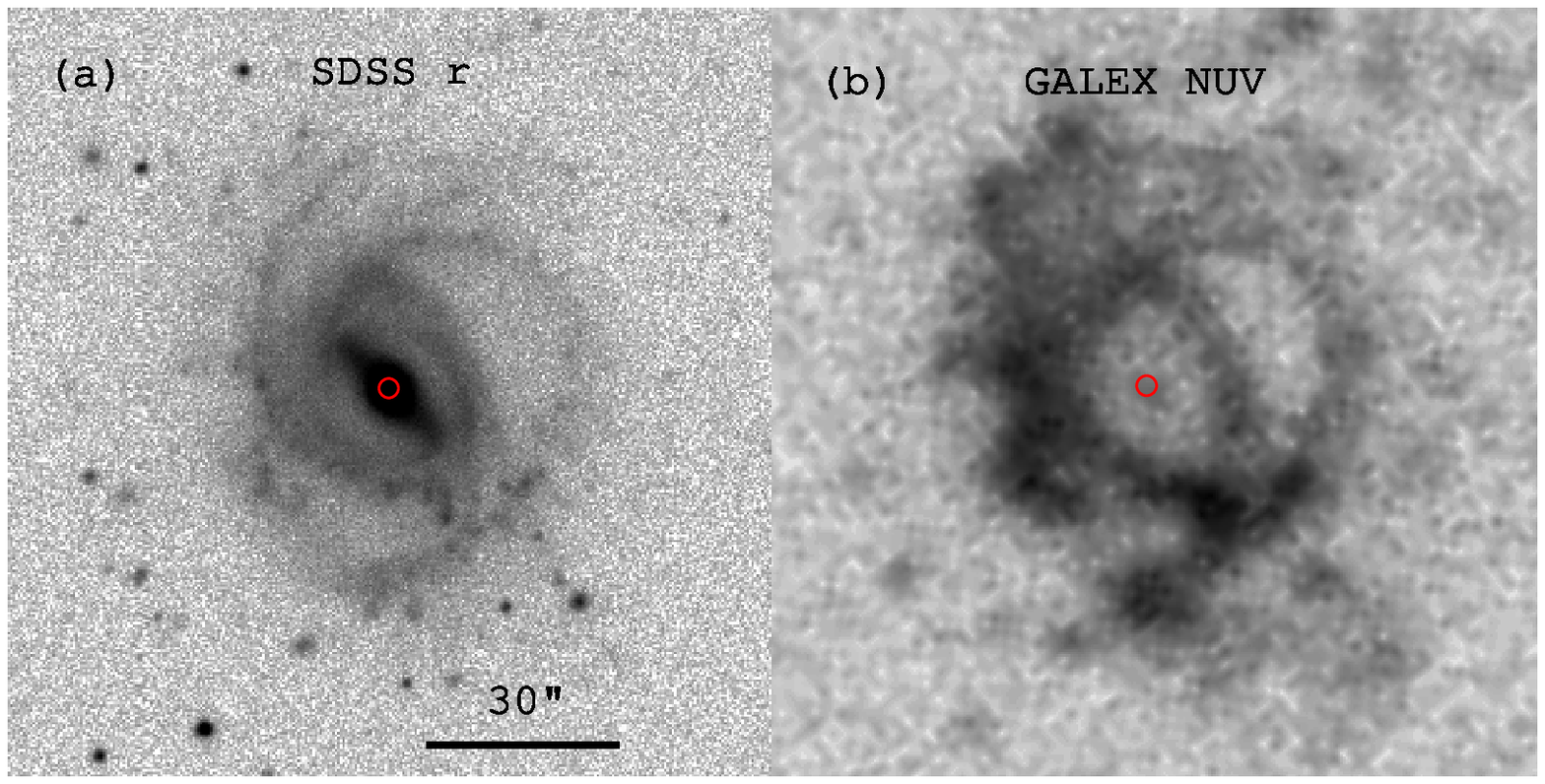}}
\centerline{\includegraphics[width=8cm]{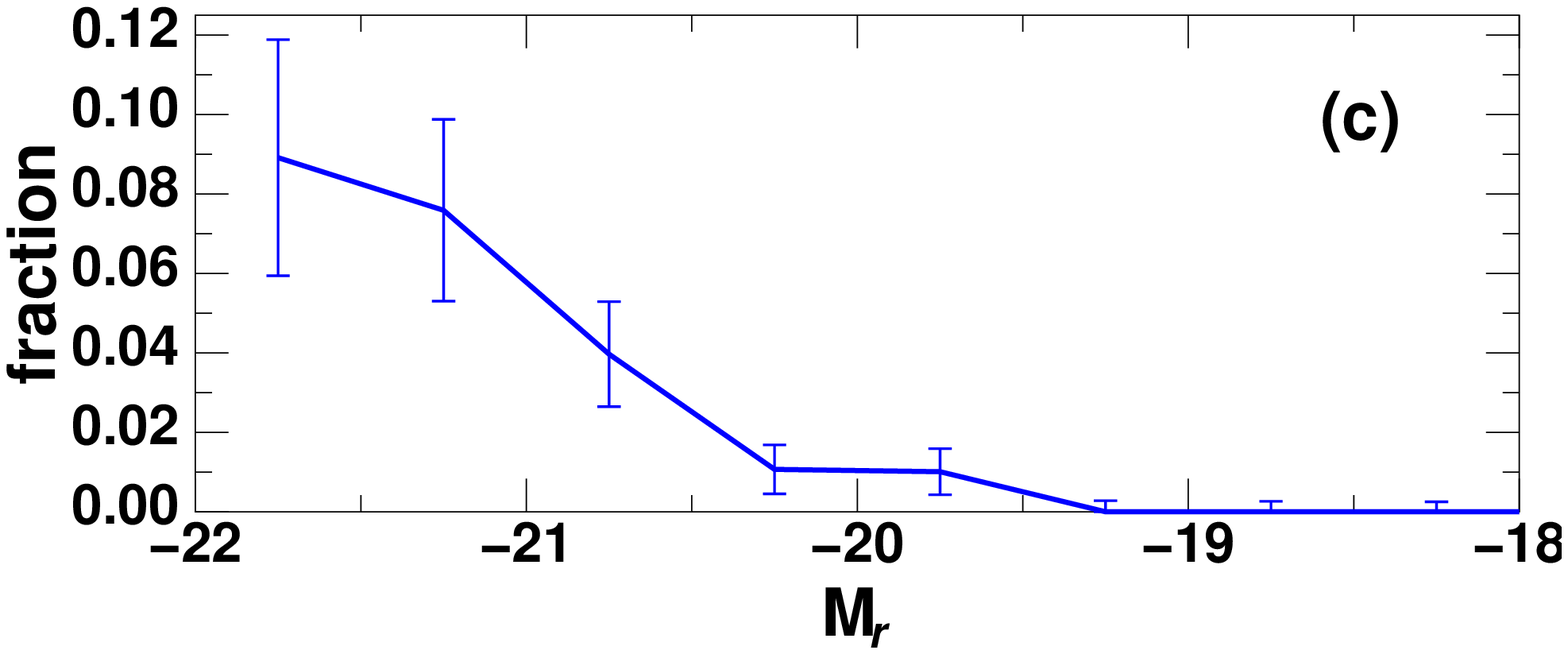}}
\caption{$r$-band (a) and NUV-band (b) images of a bright face-on
  early type spiral from our sample, highlighting the effects of the
  fibre apertures
  on classifying galaxies as passive from their SDSS spectra. The red
  circles indicate the size of the 3$''$ diameter apertures of the
  SDSS spectroscopic fibres. (c) The fraction of galaxies as a
  function of absolute magnitude 
  that are classified as passive from their SDSS spectra \mbox{(EW[H$\alpha]<2$\AA)}
  and also have blue UV-optical integrated colours (\mbox{${\rm NUV}-r<4$})
  indicative of recent star-formation.}
\label{galaxy}
\end{figure}

In Papers I and II we 
are primarily interested in the simple classification of galaxies into
passive and star-forming, and so the main issue is the number of
early-type spiral galaxies which may appear passive from spectra that
sample only their bulge, but have also normal star-forming disks.
As demonstrated earlier, the integrated \mbox{${\rm NUV}-r$} colour gives a
robust separation of passive and star-forming galaxies. Hence we can
quantify the level of aperture bias as the fraction of galaxies that
are classified as ``passive'' from their spectra by having \mbox{EW(H$\alpha)<2$\AA}, yet have global
UV-optical colours indicative of star-forming galaxies, defined as those with \mbox{${\rm NUV}-r<4$}. 

In Figure~\ref{galaxy} we show the $r-$band (panel a) and
NUV-band (panel b) images of a typical misclassified galaxy from our sample
($z=0.0323$, \mbox{M$_{r}\!=-21.71$}), with the apertures used to obtain the SDSS spectra indicated by the red
circles. The galaxy is a clearly face-on early-type spiral, whose
$r$-band flux is dominated by a bulge, but also with clear spiral
arms. In the NUV-band image however, this bulge disappears almost completely,
while the extended UV-flux from star-formation in the disk and spiral
arms is now dominant. The reason for the misclassification is clear,
the SDSS aperture covers only the central bulge which is
passively-evolving as apparent from the ``hole'' in the UV-emission in
the nuclear regions, yet misses entirely the extended star-forming
regions from the outer disk and spiral arms.

In panel (c)  we show the fraction of  
galaxies misclassified due to aperture effects as a function of
absolute magnitude. We find that the level of misclassification due
to aperture effects is strongly luminosity dependent, dropping from
eight per cent at \mbox{M$_{r}\!<-21.0$} (20 out of 246) to zero for
\mbox{M$_{r}\!>-19.5$} galaxies. We indicate that aperture biases
are significant for \mbox{M$_{r}\!<-20$} galaxies, where seven per cent
are spectroscopically classified as passive, yet whose blue integrated
\mbox{${\rm NUV}-r$} colours are indicative of recent star-formation. As
expected many of these galaxies appear as face-on spiral galaxies with
prominent bulges, whose predominately old and passive stellar
populations dominate the flux within the SDSS spectral apertures.   
In contrast we find that none of the 1375 \mbox{M$_{r}\!>-19.5$} 
galaxies in the SDSS sample covered by GALEX NUV photometry are misclassified,
indicating that the classification of such low-luminosity galaxies
based on their H$\alpha$ emission as measured through the SDSS fibres
is robust against aperture biases. 

We can understand this luminosity dependence for the level of aperture
bias as the combination of two effects: (i) more luminous galaxies at
the same distance will have larger apparent sizes, and so the fraction
of flux covered by the SDSS fibres will be reduced; and (ii)
low-luminosity galaxies tend to be either late-type spirals or dwarf
ellipticals and hence do not have such significant radial gradients in
their star-formation rates. The luminosity function of early-type
spirals (Sa+b) for which aperture biases are by far the most important
has a Gaussian distribution centred at \mbox{M$_{r}\!=-21.7$} and width
\mbox{$\sigma\!\sim\!0.9$\,mag} \citep{delapparent}, and hence are
rare at \mbox{M$_{r}\!\ga-20$}.

\section{Discussion}
\label{discussion}

We have combined {\em GALEX} NUV photometry from the 3rd data
release with a volume-limited sample
of local SDSS DR4 galaxies to examine
the make-up and the environmental dependencies of the optical and
UV-optical colour-magnitude (C-M) diagrams. We consider the same SDSS
DR4 spectroscopic sample of 27\,753 galaxies in the redshift range
\mbox{$0.005<z<0.037$} as used in Paper II, that is \mbox{$\ga\!9$0\%}
complete to \mbox{M$_{r}\!=-18$}. From this sample, 4\,065 galaxies are also
covered by NUV photometry from the 490.1\,deg$^2$ area covered by both
the SDSS DR4 spectroscopic survey and {\em GALEX} GR3.

We find that \mbox{$\sim\!3$0\%} of red sequence galaxies in the optical C-M
diagram show
signs of ongoing star-formation from their spectra having
\mbox{EW(H$\alpha)>2$\AA}. This contamination is greatest at faint magnitudes \mbox{(M$_{r}\!>-19$)} and in field regions where as many as three-quarters of red sequence
galaxies are star-forming. 
This has important consequences for
understanding the build-up of the red and blue sequences being
interpreted as the hierarchical assembly of star-forming and
passively-evolving galaxies, as the colour of the galaxy cannot be
always reliably related to its star-formation history. 
Instead a
significant fraction (and in some cases the majority) of galaxies on
the red sequence have star-formation histories more in common with the
blue cloud population and appear red simply due to dust, the presence
of which is directly related to them having active ongoing star-formation.

The effect of the contamination of the faint end of the red sequence
by dusty star-forming galaxies is likely to become increasingly
important at higher redshifts, where the global star-formation rates
\citep{noeske,zheng}, and hence the effects of dust extinction are that
much higher. This produces a significant overestimation of the amount
of stellar mass already in the red sequence at a given epoch,
particularly at the low mass end, and an overestimation of the
look-back time by which the stellar mass is assembled in
passively-evolving galaxies. Similarly, the amount of stellar mass in
the blue sequence or star-forming galaxies will be underestimated, as
will the global amount of star-formation in the blue sequence. These
effects could bias the interpretation of studies looking at the
evolution of the optically-defined red and blue sequences with
redshift in terms of the hierarchical assembly and conversion of
accreted gas into stars in passive and
star-forming galaxies \citep[e.g.][]{bell,bundy,glazebrook,willmer}, the result
of which could be the over-estimation of the importance of dry mergers
to create the present day population of passively-evolving galaxies.

We find that instead the
\mbox{${\rm NUV}-r$} colour allows a much more
robust separation of passively-evolving and star-forming galaxies,
with a clear red sequence of passively-evolving galaxies at
\mbox{${\rm NUV}-r\ga5$} and a well separated blue sequence of
star-forming galaxies at \mbox{${\rm NUV}-r<4$}. We find that globally
only 8\% of UV-selected red sequence galaxies are star-forming,
i.e. one quarter of the contamination seen in the optical red sequence.
This robust separation allows the build-up of the UV-selected red sequence with
redshift and environment to be directly interpreted in terms of the assembly of
stellar mass in passively-evolving galaxies \citep[see
  also][]{arnouts,martin07}. 

In cluster and group environments we find
the UV-selected red sequence to be fully in place to at least \mbox{M$_{r}\!=-18$}, being well described by a single Schechter function with a faint-end slope $\alpha=-0.88\pm0.12$. 
In sharp contrast we find that in isolated
field regions the
number density of UV-optical red sequence galaxies declines rapidly at
magnitudes fainter than \mbox{M$_{r}\!\sim-19$} and is completely
truncated at \mbox{M$_{r}\!\sim-18$}. This confirms the findings of
Paper II that no passively-evolving dwarf galaxies are
found more than two virial radii from a massive halo, whether that be
a group, cluster of massive galaxy. Moreover we show that the
classification of dwarf galaxies into passively-evolving and
star-forming from their SDSS fibre-obtained spectra as used in Paper
II is robust to
aperture effects, finding that none of the 1375
\mbox{M$_{r}\!>-19.5$} galaxies with integrated \mbox{${\rm NUV}-r<4$} colours
indicative of recent star-formation were also spectroscopically classified
as passive with \mbox{EW(H$\alpha)<2\,$\AA}. Hence in isolated field
regions (where the majority of galaxies are found) we find that the
build-up of the red sequence of passively-evolving galaxies is
incomplete, forming from the top down and being truncated at
\mbox{M$_{r}\!\sim-18$}. In these regions only {\em internal} processes such as
merging, supernovae and AGN feedback mechanisms, and gas exhaustion
due to star-formation can be responsible for completely stopping star-formation
in galaxies, and hence cannot be effective in low-mass galaxies
\citep[for a discussion see][]{paper2} as otherwise passively-evolving
dwarf galaxies would be ubiquitous. Instead the passively-evolving
dwarf galaxies that dominate (in terms of numbers) groups and clusters
and make up the faint-end of the cluster red sequence must have had their star-formation quenched through processes directly
related to their environment such as ram-pressure stripping or tidal shocks.

\citet{arnouts} have followed the build-up with redshift of the
passive and star-forming galaxy populations as selected from their
rest-frame \mbox{${\rm NUV}-r$} colour and/or their spectral energy
  distributions obtained by fitting their $UBVRI+JK+{\rm IRAC}$
  optical+IR photometry. They are able to obtain a similarly robust
  separation of rest-frame UV-selected red and blue sequence galaxies
  to at least \mbox{$z\!\sim\!1.2$} and probably as early as
  $z=2$. For each redshift bin (using either spectroscopic redshifts
  from the VIMOS VLT Deep Survey or photometric redshifts) they
  obtain separate stellar mass functions for the quiescent/red and active/blue
  galaxy populations. They find that the stellar mass density of red
  sequence galaxies has increased by a factor two since \mbox{$z\sim1.2$},
  and that this increase can be accounted for entirely by the shutdown
  of star-formation in active galaxies without requiring additional
  growth through dry mergers, in agreement with the results
  of \citet{bell07} who derive the global star-formation
  contribution of blue and red galaxy populations from ultraviolet and
  {\em Spitzer} 24\,$\mu$m luminosities.

These results support
the downsizing paradigm whereby the red sequence is built-up from the
top down, being already largely in place at the bright end by
\mbox{$z\!\sim\!1$} \citep{bell,willmer}, and the faint end filled in at later epochs in
clusters and groups through environment-related processes such as
ram-pressure stripping or galaxy harassment. This filling in of the
faint end appears to occur mainly at \mbox{$z<1$}, and occurs earlier in the
richest clusters. \citet{tanaka07} find few faint red galaxies in four
clumps belonging to a large-scale structure at \mbox{$z\!\sim\!1.2$} suggesting
that the red sequence is truncated within groups at \mbox{M$^{*}\!+1.5$},
whereas by combining several rich clusters at \mbox{$z\!\sim\!1.2$}
\citet{strazzullo} find the red sequence to be fitted by a Schechter
function with $\alpha=-0.85$ to at least
\mbox{M$^{*}\!+2.5$}. \citet{delucia} find an increase of a factor two
  in the dwarf-to-giant ratio of red sequence galaxies from \mbox{$z\!\sim\!0.8$}
  to \mbox{$z\!\sim\!0.4$}, while \citet{stott} find evidence for a further
  doubling of the dwarf-to-giant ratio in the cluster red sequence
  since \mbox{$z\!\sim\!0.5$}. \citet{delucia} find that this
  incremental filling in of the faint cluster red sequence population
  since \mbox{$z\!\sim\!0.8$} is consistent with a simple model whereby
  blue galaxies in clusters at \mbox{$z\!\sim\!0.8$} have their
  star-formation suppressed by the hostile cluster environment.

At all epochs to \mbox{$z\!\sim\!0.8$} \citet{tanaka05} find a deficit of faint
\mbox{(M$_{V}\!>{\rm M}_{V}\!+1$)} red sequence galaxies in field
regions, with respect to cluster and group environments, although
their deficits are not as dramatic as presented here due to the use of
photometric redshifts and statistical subtraction methods to define
the field sample, as well as the use of rest-frame $U-V$
colour-magnitude diagrams to define the red sequence population. 
To obtain clearer results for the field dwarf galaxy population at these redshifts would require a large-scale
spectroscopic survey of galaxies to \mbox{$\sim{\rm M}^{*}\!+3$} in order to define both
the redshifts and environments of each galaxy, something that is
within reach of present surveys (e.g. DEEP2, VVDS), at least to \mbox{$z\!\sim\!0.4$}.

\section*{acknowledgements}
CPH acknowledges the financial supports provided
  through the European Community's Human Potential Program, under
  contract HPRN-CT-2002-0031 SISCO. The authors thank Gianni
  Busarello, Amata Mercurio and Francesco La Barbera for stimulating
  discussions during this project.

Funding for the SDSS and SDSS-II has been provided by the Alfred
P. Sloan Foundation, the Participating Institutions, the National
Science Foundation, the U.S. Department of Energy, the National
Aeronautics and Space Administration, the Japanese Monbukagakusho, the
Max Planck Society, and the Higher Education Funding Council for
England. The SDSS Web Site is http://www.sdss.org/. 

The SDSS is managed by the Astrophysical Research Consortium for the
Participating Institutions. The Participating Institutions are the
American Museum of Natural History, Astrophysical Institute Potsdam,
University of Basel, University of Cambridge, Case Western Reserve
University, University of Chicago, Drexel University, Fermilab, the
Institute for Advanced Study, the Japan Participation Group, Johns
Hopkins University, the Joint Institute for Nuclear Astrophysics, the
Kavli Institute for Particle Astrophysics and Cosmology, the Korean
Scientist Group, the Chinese Academy of Sciences (LAMOST), Los Alamos
National Laboratory, the Max-Planck-Institute for Astronomy (MPIA),
the Max-Planck-Institute for Astrophysics (MPA), New Mexico State
University, Ohio State University, University of Pittsburgh,
University of Portsmouth, Princeton University, the United States
Naval Observatory, and the University of Washington.

This research has made use of the NASA/IPAC Extragalactic Database
(NED) which is operated by the Jet Propulsion Laboratory, California
Institute of Technology, under contract with the National Aeronautics
and Space Administration.

GALEX ({\em Galaxy Evolution Explorer}) is a NASA Small Explorer,
launched in April 2003. We gratefully acknowledge NASA's support for
the construction, operation, and science analysis for the GALEX
mission, developed in cooperation with the Centre National d'Etudes
Spatiales of France and the Korean Ministry of Science and Technology.

\label{lastpage}
\end{document}